\documentclass[
superscriptaddress,
preprint,
amsmath,amssymb,
aps,
]{revtex4-1}

\usepackage{graphicx}

\usepackage{dcolumn}% Align table columns on decimal point
\usepackage{bm}% bold math
\usepackage{hyperref}% add hypertext capabilities
\usepackage[mathlines]{lineno}% Enable numbering of text and display math
\hypersetup{colorlinks=true,allcolors=blue}

\begin{document}  

%%% This one is mandatory:
\title{Surface tension of the horizon}
%%% this one is not:
% \subtitle{subtitle}
%%% In two-column articles the sequence of author names and affiliationes in the
%%% article title page is identical to the sequence of the respective \author
%%% and \affiliation macros here.  They don't need to be intertwined as in 
%%% three-column mode, though.  The correspondence between author names and 
%%% (probably multiple) affiliations is shown using numerical tags in \inst 
%%% macros and the optional arguments of the \affiliation declarations, 
%%% respectively:
\author{Liangsuo Shu}
\email{liangsuo\_shu@hust.edu.cn}
\affiliation{School of Physics, Huazhong University of Science \& Technology. Wuhan, China}
\affiliation{School of Energy and Power Engineering, Huazhong University of Science \& Technology. Wuhan, China}

\author{Kaifeng Cui}
\email{cuikaifeng@wipm.ac.cn}
\affiliation{Key Laboratory of Atom Frequency Standards, Wuhan Institute of Physics and Mathematics, Chinese Academy of Sciences, Wuhan, China}
\affiliation{University of Chinese Academy of Sciences, Beijing, China}

\author{Xiaokang Liu}
\email{xk\_liu@hust.edu.cn}
\affiliation{School of Energy and Power Engineering, Huazhong University of Science \& Technology. Wuhan, China}

\author{Zhichun Liu}
\email{zcliu@hust.edu.cn}
\affiliation{School of Energy and Power Engineering, Huazhong University of Science \& Technology. Wuhan, China}

\author{Wei Liu }
%\altaffiliation{Corresponding author.}
\email{w\_liu@hust.edu.cn}
\affiliation{School of Energy and Power Engineering, Huazhong University of Science \& Technology. Wuhan, China}

% \affiliation[1]{School of Physics, Huazhong University of Science \& Technology. Wuhan, China}
% \affiliation[2]{School of Energy and Power Engineering, Huazhong University of Science \& Technology. Wuhan, China}
% \affiliation[3]{Key Laboratory of Atom Frequency Standards, Wuhan Institute of Physics and Mathematics, Chinese Academy of Sciences, Wuhan, China}
% \affiliation[4]{University of Chinese Academy of Sciences, Beijing, China}
%%% Shortened author names for the column titles. This is necessary only if
%%% there are more than two authors.
% \shortauthors{L.Shu,K.Cui,X.Liu,Z.Liu,W.Liu}
%%% In two-column mode, abstracts are typeset inside a colored box. If the
%%% abstract text fits nicely in one column it should be typeset that way. This
%%% is achieved by giving the \shortabstract directive. If your abstract won't
%%% fit (or if you are in doubt if it will once the final typefaces are applied)
%%% please leave that decision to the editor.

\begin{abstract}
  The idea of treating the horizon of a black hole as a stretched membrane with surface tension has a long history. In this work, we discuss the microscopic origin of the surface tension of the horizon in quantum pictures of spaces, which are Bose-Einstein condensates of gravitons. The horizon is a phase interface of gravitons, the surface tension of which is found to be a result of the difference in the strength of the interaction between the gravitons on its two sides. The gravitational source, such as a Schwarzschild black hole, creates a transitional zone by changing the energy and distribution of its surrounding gravitons. Archimedes' principle for gravity can be expressed as follows: ``the gravity on an object is equal to the weight of the gravitons that it displaces.''
\end{abstract}
% \shortabstract
%%% Here, the document begins.

\maketitle
%%% Use this if the article text won't start with a \section:
% \noindent
%%% Being based on LaTeX's article class, and2012 supports the respective 
%%% sectioning level from \section to \subparagraph.

\section{Introduction}
%\subsection{subsection}% Maybe ...
Recently, Dvali and Gomez \cite{dvali_black_2013-1,dvali_landauginzburg_2012,dvali_black_2013,dvali_quantum_2014,dvali_black_2014} have developed quantum pictures of gravitational backgrounds, such as black holes, Anti-de Sitter space (AdS), de Sitter space (dS) and inflationary universes, which can be viewed as Bose-Einstein condensates (BEC) composed of N soft constituent gravitons. Some related ideas appeared in \cite{dvali_scrambling_2013,binetruy_vacuum_2012,chapline_quantum_2001,chapline_quantum_2003,berkhahn_microscopic_2013,veneziano_quantum_2013,flassig_black_2013,casadio_quantum_2013,mueck_counting_2013,brustein_origin_2014,casadio_horizon_2014,kuhnel_high-energy_2014,kuhnel_decay_2014}.

In Dvali-Gomez's quantum picture\cite{dvali_quantum_2014}, the wave-length of gravitons set by the characteristic classical size R (i.e., the curvature radius) of the system and a Schwarzschild-Anti-de Sitter space (S-AdS) was treated as a mixture of two graviton-BECs. The thermodynamics of dS, AdS, and black holes in asymptotically dS and AdS have been studied by Hawking and his coworkers, Gibbons\cite{gibbons_cosmological_1977} and Page\cite{hawking_thermodynamics_1983}. If Dvali and Gomez's quantum pictures for space have universality, S-dS space should also be a mixture of graviton-BECs. Using BEC picture of S-dS Space, Cadoni and coworkers\cite{cadoni_effective_2018,cadoni_emergence_2018} discussed dark matter, which was  explained as a result of the reaction of the dark energy fluid to the presence of baryonic matter.

In condensed matter physics, the mixtures of BECs also draw much consideration. In this field, there have been some important achievements in both  experiment\cite{myatt_production_1997,hall_measurements_1998,stenger_spin_1998,miesner_observation_1999,stamper-kurn_quantum_1999,modugno_two_2002} and the theory\cite{ho_binary_1996,timmermans_phase_1998,van_schaeybroeck_interface_2008}. According to the classification by Timmermans\cite{timmermans_phase_1998}, the separations in the mixtures of BECs can be divided into two types: phase separation and potential separation. The idea that treating the horizon of a black hole as a stretched membrane with surface tension has a long history\cite{macdonald_membrane_1985,price_membrane_1986,price_membrane_1988,huang_thermodynamical_1993} (the origin of this idea  dates back to the 1970s \cite{hanni_lines_1973,wald_black_1974}). If the separation in S-dS space is phase separation, which means that the black hole horizon is a phase separation of gravitons, then the microscopic origin of the surface tension of horizon will become easier to understand: in two phases separated by the horizon, the interaction between gravitons is different,  therefore resulting in the surface tension of the horizon. 

In a binary BEC system with phase separation, there is Laplace pressure across the interface\cite{van_schaeybroeck_interface_2008}. What is the physical meaning of the Laplace pressure across the horizon? When re-examining Verlinde's entropic gravity\cite{verlinde_origin_2011} from the perspective of quantum pictures of space, we found that the Laplace pressure across the stretched horizon can be related to its surface gravity. Outside a black hole in asymptotically (A)dS or asymptotically flat space, there is a transitional zone. Under the influence of the gravitational potential of the black hole, the energy of the gravitons in the transitional zone is changed to varying degrees depending on their distances to the horizon.The gravitons of the same temperature constitute a holographic screen (including the horizon). Displacing some gravitons, an object on the horizon expresses a buoyancy from the gravitons, which is the gravity acting on it. 

The transitional zone can also be  regarded as a development of Dvali and Gomez's thought about the wavelength of graviton to a general case: the energy of gravitons set by the curvature radius of their local space. In this way, the thermodynamic description based on quantum space provides a microscopic dynamics mechanism for the geometric description in general relativity.

\section{Pressure and thermodynamic volume of a black hole}

Since the cosmological constant has an associated pressure, the thermodynamics of black hole with pressure and its conjugate thermodynamic variable, thermodynamic volume, have been discussed in\cite{kastor_enthalpy_2009,dolan_cosmological_2011,cvetic_black_2011,dolan_pressure_2011,kubiznak_p-v_2012}. For a Schwarzschild black hole in dS or AdS spacetime, the first law of thermodynamics will be
\begin{equation}
dU=TdS-PdV
\label{lab1}
\end{equation}
where
\begin{equation}
P=-\frac{\Lambda }{{8\pi }}
\label{lab2}
\end{equation}
is the associated pressure of cosmological constant $\Lambda$ and
\begin{equation}
{V_B} = \frac{{4\pi r_0^3}}{3}
\label{lab3}
\end{equation}
is the conjugate thermodynamic variable of $P$, the thermodynamic volume of the black hole with a radius of $r_0$. $c = \hbar  = G = {k_B} =1$ is used in this work.

$V_B$ is identified as the volume excluded by the black hole horizon from a spatial slice exterior to the black hole\cite{kastor_enthalpy_2009}. In Dvali and Gomez's quantum pictures of space \cite{dvali_black_2013-1,dvali_landauginzburg_2012,dvali_black_2013,dvali_quantum_2014,dvali_black_2014}, the thermodynamic volume of the black hole should be the volume of the internal gravitons. When $\Lambda=0$, which means $P=0$, the change in thermodynamic volume of a black hole, $dV$, does not affect its energy, 
\begin{equation}
\begin{array}{*{20}{c}}
{dU=TdS}&{\ \ (\Lambda  = 0}
\end{array})
\label{lab4}
\end{equation}

\section{Surface gravity and surface tension}

\begin{figure}[ht]
	\centering
	\includegraphics[width=0.6\textwidth]{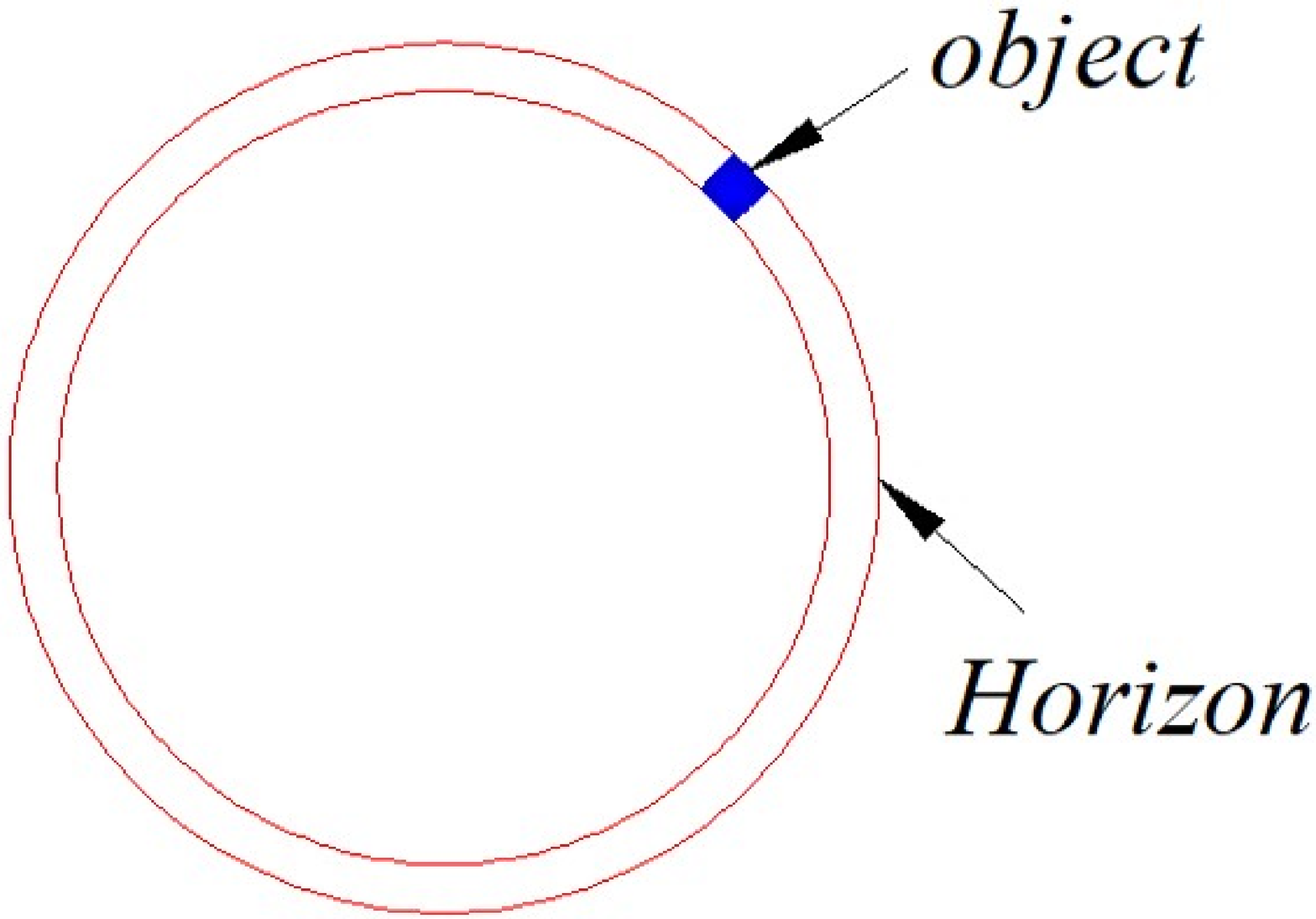}
	\caption[Object merges with the microscopic degrees of freedom on the horizon]{
		The object contributes a part of area of the horizon. The Laplace pressure across the horizon caused by its surface tension will make the object to express a gravity.}
	\label{fig:label}
\end{figure}

As aforementioned in the introduction, the internal and external gravitons of a black hole form a binary BEC system. Hereinafter, for convenience of description, the internal and external gravitons are respectively referred as gravitons in State I and State II. In a binary BEC system with phase separation, there is Laplace pressure across the interface \cite{van_schaeybroeck_interface_2008}. Therefore, for the horizon as an interface with the surface tension of ${{\sigma _H}}$, the difference in pressure across it is
\begin{equation}
\Delta P = P-{P_{out}} = \frac{{2{\sigma _H}}}{{{r_0}}}
\label{lab5}
\end{equation}
where $P_{out}$ is the gravitational pressures outside the horizon. When re-examining Verlinde's entropic gravity \cite{verlinde_origin_2011} from the quantum pictures of space perspective, we found that the pressure difference described by equation (\ref{lab5}) can be linked with the surface gravity of the horizon. Verlinde's derivation can briefly be reviewed as described below.

Motivated by Bekenstein's argument leading to his area law for the black hole entropy, that when a particle is one Compton wavelength from the horizon, it is considered to be part of the black hole and brings in 1 bit of information\cite{bekenstein_black_1973}, Verlinde provided his thermodynamic interpretations of gravity based on the holographic principle\cite{hooft_dimensional_1993,susskind_world_1995}. 

The change in entropy associated with the information on the boundary equals
\begin{equation}
\begin{array}{*{20}{c}}
{\Delta S = 2\pi }&{\ \ when\ \ }&{\Delta r = 1/m}
\end{array}
\label{lab6}
\end{equation}
The holographic relation
\begin{equation}
N = A
\label{lab7}
\end{equation}
describes that the number of bits on the boundary of a gravitational system, $N$, is proportional to the area of boundary, $A$. In Verlinde's work, this boundary is a holographic screen, the energy of which is evenly distributed over the occupied bits, $N$ and is equivalent to the Misner-Sharp mass $M$ that would emerge in the part of space surrounded by the screen. $M$ is also the total energy of the gravitons in State I. The equipartition rule is assumed to hold for the holographic screen
\begin{equation}
E = M = \frac{1}{2}N{T_r}
\label{lab8}
\end{equation}
which describes that the energy of a gravitational system, $M$, is divided evenly over the bits $N$ at  the temperature of $T_r$. For a horizon or general holographic screen,
\begin{equation}
A = 4\pi {r^2}
\label{lab9}
\end{equation}
Therefore, $T_r$ can be determined by equations (\ref{lab8}$-$\ref{lab9}) as 
\begin{equation}
{T_r} = \frac{M}{{2\pi {r^2}}}
\label{lab10}
\end{equation}
When the object moves over a distance of one reduced Compton wavelength, the entropy change will make it experience an effective entropic force. From equations (\ref{lab6}$-$\ref{lab10}), this entropic force can be obtained as
\begin{equation}
F(r) = \frac{{{T_r}\Delta S}}{{\Delta r}} = \frac{{Mm}}{{{r^2}}}
\label{lab11}
\end{equation}
From the Novikov coordinates \cite{novikov_black_1997}, the equivalent gravitational acceleration of a free-falling object relative to a hypothetical local inertial observer in Schwarzschild metric ($\Lambda=0$) is 
\begin{equation}
g(r) = \frac{{{d^2}r}}{{d{\tau ^2}}} = \frac{M}{{{r^2}}}
\label{lab12}
\end{equation}
where $\tau $ is the proper time of the free-falling object when it reaches the coordinate $r$. Comparing equation (\ref{lab11}) and equation (\ref{lab12}), the entropic force is found to be the gravity of the object from the view of a hypothetical local inertial observer at the coordinate $r$.

In Dvali and Gomez's quantum pictures of space, a holographic screen (including the horizon) should be a layer of gravitons with the same temperature and energy. When an object crosses the horizon, $\Delta P$ across the horizon will make the object experience an effective force, 
\begin{equation}
F({r_0}) =  - \Delta PdA = \frac{{Mm}}{{r_0^2}}
\label{lab13}
\end{equation}
where $dA$ is the increase in horizon area when the object crosses the horizon and merges with the microscopic degrees of freedom on the horizon. From \cite{verlinde_origin_2011}, we know
\begin{equation}
dA = dN = \frac{{2m}}{{{T_{{r_0}}}}}
\label{lab14}
\end{equation}

From equations (\ref{lab5}), (\ref{lab13}) and (\ref{lab14}), the surface tension of a Schwarzschild black hole is
\begin{equation}
{\sigma _H} =- \frac{{\kappa _H^2{r_0 }}}{{8\pi }}=- \frac{1}{{32\pi {r_0}}}
\label{lab15}
\end{equation}
where $\kappa _H$ is the black hole surface gravity. 

The special case of $\Lambda=0$ is discussed above. Chang-Young and his co-workers \cite{chang-young_schwarzschild-sitter_2011}  have generalized Verlinde's entropic gravity to Schwarzschild-de Sitter space by dividing the space into two parts. At each part, equation (\ref{lab8}) still hold. Therefore, the above analysis method for $\Lambda=0$ can be generalized to Schwarzschild-de Sitter space.

For a Schwarzschild black hole in an asymptotically de Sitter space when $\Lambda  \ne 0$, the black hole surface gravity is
\begin{equation}
{\kappa _H} = \frac{\Lambda }{{6{r_ + }}}({r_{ +  + }} - {r_ + })({r_ + } - {r_{ -  - }})
\label{lab16}
\end{equation}
where ${r_{ +  + }}$, ${r_{ + }}$, and ${r_{ -  - }}$ are three roots of
\begin{equation}
3r - 6M - \Lambda {r^3} = 0
\label{lab17}
\end{equation}
The surface tension of the black hole horizon in a Schwarzschild-de Sitter space can be described as
\begin{equation}
{\sigma _H} =  - \frac{{\kappa _H^2{r_ + }}}{{8\pi }}
\label{lab18}
\end{equation}

In summary, it is the difference in the interaction strength between the gravitons on two sides of a horizon that results in the surface gravity and the surface tension of the horizon. The interaction between gravitons in BEC is not the focus of this work. Cadoni and coworkers\cite{cadoni_effective_2018,cadoni_emergence_2018} have worked on this topic. In \cite{cadoni_effective_2018,cadoni_emergence_2018}, the local Newtonian gravity was recovered from the cosmological condensate (and new phenomenological consequences emerge) and the dark matter was explained as an appearance of the reaction of the dark energy fluid to the presence of baryonic matter.

\section{Archimedes' principle for gravity}

As aforementioned in the previous section, a holographic screen is a layer of gravitons with the same temperature and energy in quantum pictures of space. Assuming the thickness of the holographic screen at coordinate $r$ is $l_r$, the volume of the holographic screen $V$ will be
\begin{equation}
V = A{l_r}
\label{lab19}
\end{equation}
Of the volume $V$, particle of mass $m$ on the horizon contributes
\begin{equation}
dV = dA{l_r} = \frac{m}{M}A{l_r}
\label{lab20}
\end{equation}
The energy density of the holographic screen $\rho$ is
\begin{equation}
\rho  = \frac{E}{V} = \frac{M}{{A{l_r}}}
\label{lab21}
\end{equation}
Displacing some gravitons, the particle should express a buoyancy from the gravitons. According to the famous Archimedes' principle, this buoyancy $F_b$ is equal to the weight of the gravitons displaced by the particle
\begin{equation}
{F_b} = \rho gdV
\label{lab22}
\end{equation}
where $g$ is the local gravitational acceleration. Substituting equations (\ref{lab20}) and (\ref{lab21}) into equation (\ref{lab22}), we can find that
\begin{equation}
{F_b} = mg
\label{lab23}
\end{equation}

Therefore, an object's gravity is its buoyancy in space. Archimedes' principle for gravity can be described as follows: ``the gravity on an object is equal to the weight of the gravitons displaced by it.'' 

The energy of the gravitons in State II in the transitional zone outside the black hole has two sources: its own energy and the induced energy owing to the existence of a gravitational source. In a Schwarzschild-de Sitter space, both sources exist; in an asymptotically flat space, only the induced energy exists. It should particularly be mentioned that only the first source will affect the energy of the holographic screen.

In an earlier thermodynamic interpretations of gravity \cite{jacobson_thermodynamics_1995}, Jacobson proposed a view that Einstein equation is an equation of state of space. The density of the holographic screen compositing of gravitons is a kind of vacuum energy, which therefore can be regarded as an equivalent local cosmological constant,$\Lambda \left( r \right) = 8\pi \rho \left( r \right)$, with an associated pressure of $P\left( r \right) =  - \rho \left( r \right)$. The gravitational intensity is determined by the pressure gradient in the transitional zone. In the quantum picture of space, the counterpart of the metric curvature is the temperature of the graviton.

\section{Stability of BEC black hole}
Dvali and Gomez's quantum pictures of gravitational backgrounds \cite{dvali_black_2013-1,dvali_landauginzburg_2012,dvali_black_2013,dvali_quantum_2014,dvali_black_2014}  provide an insightful perspective of space and may shed light on quantum gravity. However, the attractive potential of graviton will cause a problem:``why does the energy of a BEC black hole not collapse to its singularity like a classic black hole?'' Dvali and Gomez thought the black hole is a Bose-Einstein condensate ``frozen'' at a quantum critical point, where the system achieves a delicate balance: neither collapsing nor expanding. This appears to be a strong constraint. 
According to the works of Cadoni and coworkers\cite{cadoni_effective_2018,cadoni_emergence_2018}, we know that gravitons can be pulled out from  BEC by local baryonic sources. In this work, inspired by the thermodynamics of the gravitational system \cite{gibbons_cosmological_1977,hawking_thermodynamics_1983,verlinde_origin_2011,chang-young_schwarzschild-sitter_2011,jacobson_thermodynamics_1995}, we found that the local thermal equilibrium should be responsible for the stability of BEC black hole.

The local thermal equilibrium is an important assumption in both Verlinde's entropic gravity \cite{verlinde_origin_2011} and Jacobson's thermodynamic derivation of the Einstein equation \cite{jacobson_thermodynamics_1995}. Moreover, it is also a basic assumption in a thin film model (also referred as a membrane model) provided by Zhao and his coworkers \cite{li_entropy_2000,li_black_2001,liu_improved_2001} by improving 't Hooft's brick-wall model \cite{t_hooft_quantum_1985} in order to calculate the statistical entropies of nonstatic black holes or black holes in nonthermal equilibrium. It was found that the local equilibrium near the horizon should be maintained for the black holes with a mass far greater than the Planck scale [48]. The horizon of a BEC-black hole is a phase interface separating two different graviton-BECs. When the local equilibrium near the horizon is maintained, two graviton-BECs separated by the horizon can coexist similar to saturated water and saturated steam. 

\section{Conclusion}

The idea of treating the horizon of a black hole as a stretched membrane with surface tension has a long history. In this work, we discussed the microscopic origin of the surface tension of the horizon by combining Verlinde's entropic \cite{verlinde_origin_2011} and Dvali-Gomez quantum pictures of spaces \cite{dvali_black_2013-1,dvali_landauginzburg_2012,dvali_black_2013,dvali_quantum_2014,dvali_black_2014}. The horizon of a BEC-black hole is a phase interface separating two different graviton-BECs. The strength of the interaction between the gravitons on two sides of a horizon is different, which results in the surface gravity and the surface tension of the horizon. 

In the quantum picture of space, the dynamics of gravity can be expressed as follows:

1.	Spaces are Bose-Einstein condensates of gravitons.

2.	When any gravitational source, such as a Schwarzschild black hole, enters these spaces, it creates a transitional zone by changing the energy and distribution of its surrounding gravitons. The gravitons in the transitional zone obtain induced energy.

3.	The gravitons of the same temperature constitute holographic screens. By displacing some gravitons, an object on the horizon expresses a buoyancy from the gravitons, which is the gravity acting on it. 

\section*{Acknowledgements}

Many thanks to the anonymous reviewers for their valuable comments and suggestions, to Wei Zhong and Guoliang Xu for discussion on surface tension and thermodynamics of surface. This work is supported by the National Science Foundation of China (No. 51736004 and No.51776079).

\bibliography{prop}

\begin{thebibliography}{53}%
\makeatletter
\providecommand \@ifxundefined [1]{%
 \@ifx{#1\undefined}
}%
\providecommand \@ifnum [1]{%
 \ifnum #1\expandafter \@firstoftwo
 \else \expandafter \@secondoftwo
 \fi
}%
\providecommand \@ifx [1]{%
 \ifx #1\expandafter \@firstoftwo
 \else \expandafter \@secondoftwo
 \fi
}%
\providecommand \natexlab [1]{#1}%
\providecommand \enquote  [1]{``#1''}%
\providecommand \bibnamefont  [1]{#1}%
\providecommand \bibfnamefont [1]{#1}%
\providecommand \citenamefont [1]{#1}%
\providecommand \href@noop [0]{\@secondoftwo}%
\providecommand \href [0]{\begingroup \@sanitize@url \@href}%
\providecommand \@href[1]{\@@startlink{#1}\@@href}%
\providecommand \@@href[1]{\endgroup#1\@@endlink}%
\providecommand \@sanitize@url [0]{\catcode `\\12\catcode `\$12\catcode
  `\&12\catcode `\#12\catcode `\^12\catcode `\_12\catcode `\%12\relax}%
\providecommand \@@startlink[1]{}%
\providecommand \@@endlink[0]{}%
\providecommand \url  [0]{\begingroup\@sanitize@url \@url }%
\providecommand \@url [1]{\endgroup\@href {#1}{\urlprefix }}%
\providecommand \urlprefix  [0]{URL }%
\providecommand \Eprint [0]{\href }%
\providecommand \doibase [0]{http://dx.doi.org/}%
\providecommand \selectlanguage [0]{\@gobble}%
\providecommand \bibinfo  [0]{\@secondoftwo}%
\providecommand \bibfield  [0]{\@secondoftwo}%
\providecommand \translation [1]{[#1]}%
\providecommand \BibitemOpen [0]{}%
\providecommand \bibitemStop [0]{}%
\providecommand \bibitemNoStop [0]{.\EOS\space}%
\providecommand \EOS [0]{\spacefactor3000\relax}%
\providecommand \BibitemShut  [1]{\csname bibitem#1\endcsname}%
\let\auto@bib@innerbib\@empty
%</preamble>
\bibitem [{\citenamefont {Dvali}\ and\ \citenamefont
  {Gomez}(2013{\natexlab{a}})}]{dvali_black_2013-1}%
  \BibitemOpen
  \bibfield  {author} {\bibinfo {author} {\bibfnamefont {G.}~\bibnamefont
  {Dvali}}\ and\ \bibinfo {author} {\bibfnamefont {C.}~\bibnamefont {Gomez}},\
  }\href {http://onlinelibrary.wiley.com/doi/10.1002/prop.201300001/pdf}
  {\bibfield  {journal} {\bibinfo  {journal} {Fortschr. Phys.}\ }\textbf
  {\bibinfo {volume} {61}},\ \bibinfo {pages} {742} (\bibinfo {year}
  {2013}{\natexlab{a}})}\BibitemShut {NoStop}%
\bibitem [{\citenamefont {Dvali}\ and\ \citenamefont
  {Gomez}(2012)}]{dvali_landauginzburg_2012}%
  \BibitemOpen
  \bibfield  {author} {\bibinfo {author} {\bibfnamefont {G.}~\bibnamefont
  {Dvali}}\ and\ \bibinfo {author} {\bibfnamefont {C.}~\bibnamefont {Gomez}},\
  }\href {http://www.sciencedirect.com/science/article/pii/S0370269312008568}
  {\bibfield  {journal} {\bibinfo  {journal} {Phys. Lett. B}\ }\textbf
  {\bibinfo {volume} {716}},\ \bibinfo {pages} {240} (\bibinfo {year}
  {2012})}\BibitemShut {NoStop}%
\bibitem [{\citenamefont {Dvali}\ and\ \citenamefont
  {Gomez}(2013{\natexlab{b}})}]{dvali_black_2013}%
  \BibitemOpen
  \bibfield  {author} {\bibinfo {author} {\bibfnamefont {G.}~\bibnamefont
  {Dvali}}\ and\ \bibinfo {author} {\bibfnamefont {C.}~\bibnamefont {Gomez}},\
  }\href {http://www.oalib.com/paper/3389714} {\bibfield  {journal} {\bibinfo
  {journal} {Phys. Lett. B}\ }\textbf {\bibinfo {volume} {719}},\ \bibinfo
  {pages} {419} (\bibinfo {year} {2013}{\natexlab{b}})}\BibitemShut {NoStop}%
\bibitem [{\citenamefont {Dvali}\ and\ \citenamefont
  {Gomez}(2014{\natexlab{a}})}]{dvali_quantum_2014}%
  \BibitemOpen
  \bibfield  {author} {\bibinfo {author} {\bibfnamefont {G.}~\bibnamefont
  {Dvali}}\ and\ \bibinfo {author} {\bibfnamefont {C.}~\bibnamefont {Gomez}},\
  }\href {\doibase 10.1088/1475-7516/2014/01/023} {\bibfield  {journal}
  {\bibinfo  {journal} {J. Cosmol. Astropart. P}\ }\textbf {\bibinfo {volume}
  {2014}},\ \bibinfo {pages} {023} (\bibinfo {year}
  {2014}{\natexlab{a}})}\BibitemShut {NoStop}%
\bibitem [{\citenamefont {Dvali}\ and\ \citenamefont
  {Gomez}(2014{\natexlab{b}})}]{dvali_black_2014}%
  \BibitemOpen
  \bibfield  {author} {\bibinfo {author} {\bibfnamefont {G.}~\bibnamefont
  {Dvali}}\ and\ \bibinfo {author} {\bibfnamefont {C.}~\bibnamefont {Gomez}},\
  }\href {\doibase 10.1140/epjc/s10052-014-2752-3} {\bibfield  {journal}
  {\bibinfo  {journal} {Eur. Phys. J. C}\ }\textbf {\bibinfo {volume} {74}}
  (\bibinfo {year} {2014}{\natexlab{b}}),\
  10.1140/epjc/s10052-014-2752-3}\BibitemShut {NoStop}%
\bibitem [{\citenamefont {Dvali}\ \emph {et~al.}(2013)\citenamefont {Dvali},
  \citenamefont {Flassig}, \citenamefont {Gomez}, \citenamefont {Pritzel},\
  and\ \citenamefont {Wintergerst}}]{dvali_scrambling_2013}%
  \BibitemOpen
  \bibfield  {author} {\bibinfo {author} {\bibfnamefont {G.}~\bibnamefont
  {Dvali}}, \bibinfo {author} {\bibfnamefont {D.}~\bibnamefont {Flassig}},
  \bibinfo {author} {\bibfnamefont {C.}~\bibnamefont {Gomez}}, \bibinfo
  {author} {\bibfnamefont {A.}~\bibnamefont {Pritzel}}, \ and\ \bibinfo
  {author} {\bibfnamefont {N.}~\bibnamefont {Wintergerst}},\ }\href {\doibase
  10.1103/PhysRevD.88.124041} {\bibfield  {journal} {\bibinfo  {journal} {Phys.
  Rev. D}\ }\textbf {\bibinfo {volume} {88}} (\bibinfo {year} {2013}),\
  10.1103/PhysRevD.88.124041}\BibitemShut {NoStop}%
\bibitem [{\citenamefont {Binetruy}(2012)}]{binetruy_vacuum_2012}%
  \BibitemOpen
  \bibfield  {author} {\bibinfo {author} {\bibfnamefont {P.}~\bibnamefont
  {Binetruy}},\ }\href {http://arxiv.org/abs/1208.4645} {\bibfield  {journal}
  {\bibinfo  {journal} {arXiv:1208.4645 [gr-qc, physics:hep-th]}\ } (\bibinfo
  {year} {2012})},\ \bibinfo {note} {arXiv: 1208.4645}\BibitemShut {NoStop}%
\bibitem [{\citenamefont {Chapline}\ \emph {et~al.}(2001)\citenamefont
  {Chapline}, \citenamefont {Hohlfeld}, \citenamefont {Laughlin},\ and\
  \citenamefont {Santiago}}]{chapline_quantum_2001}%
  \BibitemOpen
  \bibfield  {author} {\bibinfo {author} {\bibfnamefont {G.}~\bibnamefont
  {Chapline}}, \bibinfo {author} {\bibfnamefont {E.}~\bibnamefont {Hohlfeld}},
  \bibinfo {author} {\bibfnamefont {R.~B.}\ \bibnamefont {Laughlin}}, \ and\
  \bibinfo {author} {\bibfnamefont {D.~I.}\ \bibnamefont {Santiago}},\ }\href
  {\doibase 10.1080/13642810108221981} {\bibfield  {journal} {\bibinfo
  {journal} {Phil. Mag. B}\ }\textbf {\bibinfo {volume} {81}},\ \bibinfo
  {pages} {235} (\bibinfo {year} {2001})}\BibitemShut {NoStop}%
\bibitem [{\citenamefont {Chapline}(2003)}]{chapline_quantum_2003}%
  \BibitemOpen
  \bibfield  {author} {\bibinfo {author} {\bibfnamefont {G.}~\bibnamefont
  {Chapline}},\ }\href {\doibase 10.1142/S0217751X03016380} {\bibfield
  {journal} {\bibinfo  {journal} {Int. J. Mod. Phys. A}\ }\textbf {\bibinfo
  {volume} {18}},\ \bibinfo {pages} {3587} (\bibinfo {year}
  {2003})}\BibitemShut {NoStop}%
\bibitem [{\citenamefont {Berkhahn}\ \emph {et~al.}(2013)\citenamefont
  {Berkhahn}, \citenamefont {Müller}, \citenamefont {Niedermann},\ and\
  \citenamefont {Schneider}}]{berkhahn_microscopic_2013}%
  \BibitemOpen
  \bibfield  {author} {\bibinfo {author} {\bibfnamefont {F.}~\bibnamefont
  {Berkhahn}}, \bibinfo {author} {\bibfnamefont {S.}~\bibnamefont {Müller}},
  \bibinfo {author} {\bibfnamefont {F.}~\bibnamefont {Niedermann}}, \ and\
  \bibinfo {author} {\bibfnamefont {R.}~\bibnamefont {Schneider}},\ }\href
  {\doibase 10.1088/1475-7516/2013/08/028} {\bibfield  {journal} {\bibinfo
  {journal} {J. Cosmol. Astropart. Phys.}\ }\textbf {\bibinfo {volume}
  {2013}},\ \bibinfo {pages} {028} (\bibinfo {year} {2013})}\BibitemShut
  {NoStop}%
\bibitem [{\citenamefont {Veneziano}(2013)}]{veneziano_quantum_2013}%
  \BibitemOpen
  \bibfield  {author} {\bibinfo {author} {\bibfnamefont {G.}~\bibnamefont
  {Veneziano}},\ }\href {\doibase 10.1088/0264-9381/30/9/092001} {\bibfield
  {journal} {\bibinfo  {journal} {Class. Quantum Grav.}\ }\textbf {\bibinfo
  {volume} {30}},\ \bibinfo {pages} {092001} (\bibinfo {year}
  {2013})}\BibitemShut {NoStop}%
\bibitem [{\citenamefont {Flassig}\ \emph {et~al.}(2013)\citenamefont
  {Flassig}, \citenamefont {Pritzel},\ and\ \citenamefont
  {Wintergerst}}]{flassig_black_2013}%
  \BibitemOpen
  \bibfield  {author} {\bibinfo {author} {\bibfnamefont {D.}~\bibnamefont
  {Flassig}}, \bibinfo {author} {\bibfnamefont {A.}~\bibnamefont {Pritzel}}, \
  and\ \bibinfo {author} {\bibfnamefont {N.}~\bibnamefont {Wintergerst}},\
  }\href {\doibase 10.1103/PhysRevD.87.084007} {\bibfield  {journal} {\bibinfo
  {journal} {Phys. Rev. D}\ }\textbf {\bibinfo {volume} {87}} (\bibinfo {year}
  {2013}),\ 10.1103/PhysRevD.87.084007}\BibitemShut {NoStop}%
\bibitem [{\citenamefont {Casadio}\ and\ \citenamefont
  {Orlandi}(2013)}]{casadio_quantum_2013}%
  \BibitemOpen
  \bibfield  {author} {\bibinfo {author} {\bibfnamefont {R.}~\bibnamefont
  {Casadio}}\ and\ \bibinfo {author} {\bibfnamefont {A.}~\bibnamefont
  {Orlandi}},\ }\href {\doibase 10.1007/JHEP08(2013)025} {\bibfield  {journal}
  {\bibinfo  {journal} {J. High Energy Phys.}\ }\textbf {\bibinfo {volume}
  {2013}} (\bibinfo {year} {2013}),\ 10.1007/JHEP08(2013)025},\ \bibinfo {note}
  {arXiv: 1302.7138}\BibitemShut {NoStop}%
\bibitem [{\citenamefont {Mueck}(2013)}]{mueck_counting_2013}%
  \BibitemOpen
  \bibfield  {author} {\bibinfo {author} {\bibfnamefont {W.}~\bibnamefont
  {Mueck}},\ }\href {\doibase 10.1140/epjc/s10052-013-2679-0} {\bibfield
  {journal} {\bibinfo  {journal} {Eur. Phys. J. C}\ }\textbf {\bibinfo {volume}
  {73}} (\bibinfo {year} {2013}),\ 10.1140/epjc/s10052-013-2679-0},\ \bibinfo
  {note} {arXiv: 1310.6909}\BibitemShut {NoStop}%
\bibitem [{\citenamefont {Brustein}(2014)}]{brustein_origin_2014}%
  \BibitemOpen
  \bibfield  {author} {\bibinfo {author} {\bibfnamefont {R.}~\bibnamefont
  {Brustein}},\ }\href {\doibase 10.1002/prop.201300037} {\bibfield  {journal}
  {\bibinfo  {journal} {Fortschritte Der Physik}\ }\textbf {\bibinfo {volume}
  {62}},\ \bibinfo {pages} {255} (\bibinfo {year} {2014})}\BibitemShut
  {NoStop}%
\bibitem [{\citenamefont {Casadio}\ and\ \citenamefont
  {Scardigli}(2014)}]{casadio_horizon_2014}%
  \BibitemOpen
  \bibfield  {author} {\bibinfo {author} {\bibfnamefont {R.}~\bibnamefont
  {Casadio}}\ and\ \bibinfo {author} {\bibfnamefont {F.}~\bibnamefont
  {Scardigli}},\ }\href {\doibase 10.1140/epjc/s10052-013-2685-2} {\bibfield
  {journal} {\bibinfo  {journal} {Eur. Phys. J. C}\ }\textbf {\bibinfo {volume}
  {74}},\ \bibinfo {pages} {2685} (\bibinfo {year} {2014})}\BibitemShut
  {NoStop}%
\bibitem [{\citenamefont {Kühnel}\ and\ \citenamefont
  {Bo}(2014)}]{kuhnel_high-energy_2014}%
  \BibitemOpen
  \bibfield  {author} {\bibinfo {author} {\bibfnamefont {F.}~\bibnamefont
  {Kühnel}}\ and\ \bibinfo {author} {\bibfnamefont {S.}~\bibnamefont {Bo}},\
  }\href {http://link.springer.com/article/10.1007/JHEP12(2014)016} {\bibfield
  {journal} {\bibinfo  {journal} {J. High Energy Phys.}\ }\textbf {\bibinfo
  {volume} {2014}},\ \bibinfo {pages} {1} (\bibinfo {year} {2014})}\BibitemShut
  {NoStop}%
\bibitem [{\citenamefont {Kühnel}\ and\ \citenamefont
  {Sundborg}(2014)}]{kuhnel_decay_2014}%
  \BibitemOpen
  \bibfield  {author} {\bibinfo {author} {\bibfnamefont {F.}~\bibnamefont
  {Kühnel}}\ and\ \bibinfo {author} {\bibfnamefont {B.}~\bibnamefont
  {Sundborg}},\ }\href {\doibase 10.1103/PhysRevD.90.064025} {\bibfield
  {journal} {\bibinfo  {journal} {Phys. Rev. D}\ }\textbf {\bibinfo {volume}
  {90}} (\bibinfo {year} {2014}),\ 10.1103/PhysRevD.90.064025}\BibitemShut
  {NoStop}%
\bibitem [{\citenamefont {Gibbons}\ and\ \citenamefont
  {Hawking}(1977)}]{gibbons_cosmological_1977}%
  \BibitemOpen
  \bibfield  {author} {\bibinfo {author} {\bibfnamefont {G.~W.}\ \bibnamefont
  {Gibbons}}\ and\ \bibinfo {author} {\bibfnamefont {S.~W.}\ \bibnamefont
  {Hawking}},\ }\href {\doibase 10.1103/PhysRevD.15.2738} {\bibfield  {journal}
  {\bibinfo  {journal} {Phys. Rev. D}\ }\textbf {\bibinfo {volume} {15}},\
  \bibinfo {pages} {2738} (\bibinfo {year} {1977})}\BibitemShut {NoStop}%
\bibitem [{\citenamefont {Hawking}\ and\ \citenamefont
  {Page}(1983)}]{hawking_thermodynamics_1983}%
  \BibitemOpen
  \bibfield  {author} {\bibinfo {author} {\bibfnamefont {S.~W.}\ \bibnamefont
  {Hawking}}\ and\ \bibinfo {author} {\bibfnamefont {D.~N.}\ \bibnamefont
  {Page}},\ }\href {\doibase 10.1007/BF01208266} {\bibfield  {journal}
  {\bibinfo  {journal} {Commun. Math. Phys.}\ }\textbf {\bibinfo {volume}
  {87}},\ \bibinfo {pages} {577} (\bibinfo {year} {1983})}\BibitemShut
  {NoStop}%
\bibitem [{\citenamefont {Cadoni}\ \emph
  {et~al.}(2018{\natexlab{a}})\citenamefont {Cadoni}, \citenamefont {Casadio},
  \citenamefont {Giusti}, \citenamefont {Mück},\ and\ \citenamefont
  {Tuveri}}]{cadoni_effective_2018}%
  \BibitemOpen
  \bibfield  {author} {\bibinfo {author} {\bibfnamefont {M.}~\bibnamefont
  {Cadoni}}, \bibinfo {author} {\bibfnamefont {R.}~\bibnamefont {Casadio}},
  \bibinfo {author} {\bibfnamefont {A.}~\bibnamefont {Giusti}}, \bibinfo
  {author} {\bibfnamefont {W.}~\bibnamefont {Mück}}, \ and\ \bibinfo {author}
  {\bibfnamefont {M.}~\bibnamefont {Tuveri}},\ }\href {\doibase
  10.1016/j.physletb.2017.11.058} {\bibfield  {journal} {\bibinfo  {journal}
  {Phys. Lett. B}\ }\textbf {\bibinfo {volume} {776}},\ \bibinfo {pages} {242}
  (\bibinfo {year} {2018}{\natexlab{a}})}\BibitemShut {NoStop}%
\bibitem [{\citenamefont {Cadoni}\ \emph
  {et~al.}(2018{\natexlab{b}})\citenamefont {Cadoni}, \citenamefont {Casadio},
  \citenamefont {Giusti},\ and\ \citenamefont
  {Tuveri}}]{cadoni_emergence_2018}%
  \BibitemOpen
  \bibfield  {author} {\bibinfo {author} {\bibfnamefont {M.}~\bibnamefont
  {Cadoni}}, \bibinfo {author} {\bibfnamefont {R.}~\bibnamefont {Casadio}},
  \bibinfo {author} {\bibfnamefont {A.}~\bibnamefont {Giusti}}, \ and\ \bibinfo
  {author} {\bibfnamefont {M.}~\bibnamefont {Tuveri}},\ }\href {\doibase
  10.1103/PhysRevD.97.044047} {\bibfield  {journal} {\bibinfo  {journal} {Phys.
  Rev. D}\ }\textbf {\bibinfo {volume} {97}} (\bibinfo {year}
  {2018}{\natexlab{b}}),\ 10.1103/PhysRevD.97.044047}\BibitemShut {NoStop}%
\bibitem [{\citenamefont {Myatt}\ \emph {et~al.}(1997)\citenamefont {Myatt},
  \citenamefont {Burt}, \citenamefont {Ghrist}, \citenamefont {Cornell},\ and\
  \citenamefont {Wieman}}]{myatt_production_1997}%
  \BibitemOpen
  \bibfield  {author} {\bibinfo {author} {\bibfnamefont {C.~J.}\ \bibnamefont
  {Myatt}}, \bibinfo {author} {\bibfnamefont {E.~A.}\ \bibnamefont {Burt}},
  \bibinfo {author} {\bibfnamefont {R.~W.}\ \bibnamefont {Ghrist}}, \bibinfo
  {author} {\bibfnamefont {E.~A.}\ \bibnamefont {Cornell}}, \ and\ \bibinfo
  {author} {\bibfnamefont {C.~E.}\ \bibnamefont {Wieman}},\ }\href {\doibase
  10.1103/PhysRevLett.78.586} {\bibfield  {journal} {\bibinfo  {journal} {Phys.
  Rev. Lett.}\ }\textbf {\bibinfo {volume} {78}},\ \bibinfo {pages} {586}
  (\bibinfo {year} {1997})}\BibitemShut {NoStop}%
\bibitem [{\citenamefont {Hall}\ \emph {et~al.}(1998)\citenamefont {Hall},
  \citenamefont {Matthews}, \citenamefont {Wieman},\ and\ \citenamefont
  {Cornell}}]{hall_measurements_1998}%
  \BibitemOpen
  \bibfield  {author} {\bibinfo {author} {\bibfnamefont {D.~S.}\ \bibnamefont
  {Hall}}, \bibinfo {author} {\bibfnamefont {M.~R.}\ \bibnamefont {Matthews}},
  \bibinfo {author} {\bibfnamefont {C.~E.}\ \bibnamefont {Wieman}}, \ and\
  \bibinfo {author} {\bibfnamefont {E.~A.}\ \bibnamefont {Cornell}},\ }\href
  {http://dx.doi.org/10.1142/9789812813787_0072} {\bibfield  {journal}
  {\bibinfo  {journal} {Physical Review Letters}\ }\textbf {\bibinfo {volume}
  {81}},\ \bibinfo {pages} {1539} (\bibinfo {year} {1998})}\BibitemShut
  {NoStop}%
\bibitem [{\citenamefont {Stenger}\ \emph {et~al.}(1998)\citenamefont
  {Stenger}, \citenamefont {Inouye}, \citenamefont {Stamper-Kurn},
  \citenamefont {Miesner}, \citenamefont {Chikkatur},\ and\ \citenamefont
  {Ketterle}}]{stenger_spin_1998}%
  \BibitemOpen
  \bibfield  {author} {\bibinfo {author} {\bibfnamefont {J.}~\bibnamefont
  {Stenger}}, \bibinfo {author} {\bibfnamefont {S.}~\bibnamefont {Inouye}},
  \bibinfo {author} {\bibfnamefont {D.~M.}\ \bibnamefont {Stamper-Kurn}},
  \bibinfo {author} {\bibfnamefont {H.-J.}\ \bibnamefont {Miesner}}, \bibinfo
  {author} {\bibfnamefont {A.~P.}\ \bibnamefont {Chikkatur}}, \ and\ \bibinfo
  {author} {\bibfnamefont {W.}~\bibnamefont {Ketterle}},\ }\href {\doibase
  10.1038/24567} {\bibfield  {journal} {\bibinfo  {journal} {Nature}\ }\textbf
  {\bibinfo {volume} {396}},\ \bibinfo {pages} {345} (\bibinfo {year}
  {1998})}\BibitemShut {NoStop}%
\bibitem [{\citenamefont {Miesner}\ \emph {et~al.}(1999)\citenamefont
  {Miesner}, \citenamefont {Stamper-Kurn}, \citenamefont {Stenger},
  \citenamefont {Inouye}, \citenamefont {Chikkatur},\ and\ \citenamefont
  {Ketterle}}]{miesner_observation_1999}%
  \BibitemOpen
  \bibfield  {author} {\bibinfo {author} {\bibfnamefont {H.-J.}\ \bibnamefont
  {Miesner}}, \bibinfo {author} {\bibfnamefont {D.~M.}\ \bibnamefont
  {Stamper-Kurn}}, \bibinfo {author} {\bibfnamefont {J.}~\bibnamefont
  {Stenger}}, \bibinfo {author} {\bibfnamefont {S.}~\bibnamefont {Inouye}},
  \bibinfo {author} {\bibfnamefont {A.~P.}\ \bibnamefont {Chikkatur}}, \ and\
  \bibinfo {author} {\bibfnamefont {W.}~\bibnamefont {Ketterle}},\ }\href
  {\doibase 10.1103/PhysRevLett.82.2228} {\bibfield  {journal} {\bibinfo
  {journal} {Phys. Rev. Lett.}\ }\textbf {\bibinfo {volume} {82}},\ \bibinfo
  {pages} {2228} (\bibinfo {year} {1999})}\BibitemShut {NoStop}%
\bibitem [{\citenamefont {Stamper-Kurn}\ \emph {et~al.}(1999)\citenamefont
  {Stamper-Kurn}, \citenamefont {Miesner}, \citenamefont {Chikkatur},
  \citenamefont {Inouye}, \citenamefont {Stenger},\ and\ \citenamefont
  {Ketterle}}]{stamper-kurn_quantum_1999}%
  \BibitemOpen
  \bibfield  {author} {\bibinfo {author} {\bibfnamefont {D.~M.}\ \bibnamefont
  {Stamper-Kurn}}, \bibinfo {author} {\bibfnamefont {H.-J.}\ \bibnamefont
  {Miesner}}, \bibinfo {author} {\bibfnamefont {A.~P.}\ \bibnamefont
  {Chikkatur}}, \bibinfo {author} {\bibfnamefont {S.}~\bibnamefont {Inouye}},
  \bibinfo {author} {\bibfnamefont {J.}~\bibnamefont {Stenger}}, \ and\
  \bibinfo {author} {\bibfnamefont {W.}~\bibnamefont {Ketterle}},\ }\href
  {\doibase 10.1103/PhysRevLett.83.661} {\bibfield  {journal} {\bibinfo
  {journal} {Phys. Rev. Lett.}\ }\textbf {\bibinfo {volume} {83}},\ \bibinfo
  {pages} {661} (\bibinfo {year} {1999})}\BibitemShut {NoStop}%
\bibitem [{\citenamefont {Modugno}\ \emph {et~al.}(2002)\citenamefont
  {Modugno}, \citenamefont {Modugno}, \citenamefont {Riboli}, \citenamefont
  {Roati},\ and\ \citenamefont {Inguscio}}]{modugno_two_2002}%
  \BibitemOpen
  \bibfield  {author} {\bibinfo {author} {\bibfnamefont {G.}~\bibnamefont
  {Modugno}}, \bibinfo {author} {\bibfnamefont {M.}~\bibnamefont {Modugno}},
  \bibinfo {author} {\bibfnamefont {F.}~\bibnamefont {Riboli}}, \bibinfo
  {author} {\bibfnamefont {G.}~\bibnamefont {Roati}}, \ and\ \bibinfo {author}
  {\bibfnamefont {M.}~\bibnamefont {Inguscio}},\ }\href {\doibase
  10.1103/PhysRevLett.89.190404} {\bibfield  {journal} {\bibinfo  {journal}
  {Phys. Rev. Lett.}\ }\textbf {\bibinfo {volume} {89}},\ \bibinfo {pages}
  {190404} (\bibinfo {year} {2002})}\BibitemShut {NoStop}%
\bibitem [{\citenamefont {Ho}\ and\ \citenamefont
  {Shenoy}(1996)}]{ho_binary_1996}%
  \BibitemOpen
  \bibfield  {author} {\bibinfo {author} {\bibfnamefont {T.-L.}\ \bibnamefont
  {Ho}}\ and\ \bibinfo {author} {\bibfnamefont {V.~B.}\ \bibnamefont
  {Shenoy}},\ }\href {\doibase 10.1103/PhysRevLett.77.3276} {\bibfield
  {journal} {\bibinfo  {journal} {Phys. Rev. Lett.}\ }\textbf {\bibinfo
  {volume} {77}},\ \bibinfo {pages} {3276} (\bibinfo {year}
  {1996})}\BibitemShut {NoStop}%
\bibitem [{\citenamefont {Timmermans}(1998)}]{timmermans_phase_1998}%
  \BibitemOpen
  \bibfield  {author} {\bibinfo {author} {\bibfnamefont {E.}~\bibnamefont
  {Timmermans}},\ }\href {\doibase 10.1103/PhysRevLett.81.5718} {\bibfield
  {journal} {\bibinfo  {journal} {Phys. Rev. Lett.}\ }\textbf {\bibinfo
  {volume} {81}},\ \bibinfo {pages} {5718} (\bibinfo {year}
  {1998})}\BibitemShut {NoStop}%
\bibitem [{\citenamefont
  {Van~Schaeybroeck}(2008)}]{van_schaeybroeck_interface_2008}%
  \BibitemOpen
  \bibfield  {author} {\bibinfo {author} {\bibfnamefont {B.}~\bibnamefont
  {Van~Schaeybroeck}},\ }\href {\doibase 10.1103/PhysRevA.78.023624} {\bibfield
   {journal} {\bibinfo  {journal} {Phys. Rev. A}\ }\textbf {\bibinfo {volume}
  {78}},\ \bibinfo {pages} {023624} (\bibinfo {year} {2008})}\BibitemShut
  {NoStop}%
\bibitem [{\citenamefont {Macdonald}\ and\ \citenamefont
  {Suen}(1985)}]{macdonald_membrane_1985}%
  \BibitemOpen
  \bibfield  {author} {\bibinfo {author} {\bibfnamefont {D.~A.}\ \bibnamefont
  {Macdonald}}\ and\ \bibinfo {author} {\bibfnamefont {W.-M.}\ \bibnamefont
  {Suen}},\ }\href {\doibase 10.1103/PhysRevD.32.848} {\bibfield  {journal}
  {\bibinfo  {journal} {Phys. Rev. D}\ }\textbf {\bibinfo {volume} {32}},\
  \bibinfo {pages} {848} (\bibinfo {year} {1985})}\BibitemShut {NoStop}%
\bibitem [{\citenamefont {Price}\ and\ \citenamefont
  {Thorne}(1986)}]{price_membrane_1986}%
  \BibitemOpen
  \bibfield  {author} {\bibinfo {author} {\bibfnamefont {R.~H.}\ \bibnamefont
  {Price}}\ and\ \bibinfo {author} {\bibfnamefont {K.~S.}\ \bibnamefont
  {Thorne}},\ }\href {\doibase 10.1103/PhysRevD.33.915} {\bibfield  {journal}
  {\bibinfo  {journal} {Phys. Rev. D}\ }\textbf {\bibinfo {volume} {33}},\
  \bibinfo {pages} {915} (\bibinfo {year} {1986})}\BibitemShut {NoStop}%
\bibitem [{\citenamefont {Price}\ and\ \citenamefont
  {Thorne}(1988)}]{price_membrane_1988}%
  \BibitemOpen
  \bibfield  {author} {\bibinfo {author} {\bibfnamefont {R.~H.}\ \bibnamefont
  {Price}}\ and\ \bibinfo {author} {\bibfnamefont {K.~S.}\ \bibnamefont
  {Thorne}},\ }\href {\doibase 10.1038/scientificamerican0488-69} {\bibfield
  {journal} {\bibinfo  {journal} {Scientific American}\ }\textbf {\bibinfo
  {volume} {258}},\ \bibinfo {pages} {69} (\bibinfo {year} {1988})}\BibitemShut
  {NoStop}%
\bibitem [{\citenamefont {Huang}\ \emph {et~al.}(1993)\citenamefont {Huang},
  \citenamefont {Liu},\ and\ \citenamefont
  {Zhao}}]{huang_thermodynamical_1993}%
  \BibitemOpen
  \bibfield  {author} {\bibinfo {author} {\bibfnamefont {C.~G.}\ \bibnamefont
  {Huang}}, \bibinfo {author} {\bibfnamefont {L.}~\bibnamefont {Liu}}, \ and\
  \bibinfo {author} {\bibfnamefont {Z.}~\bibnamefont {Zhao}},\ }\href {\doibase
  10.1007/BF00759032} {\bibfield  {journal} {\bibinfo  {journal} {Gen. Rel.
  Gravit.}\ }\textbf {\bibinfo {volume} {25}},\ \bibinfo {pages} {1267}
  (\bibinfo {year} {1993})}\BibitemShut {NoStop}%
\bibitem [{\citenamefont {Hanni}\ and\ \citenamefont
  {Ruffini}(1973)}]{hanni_lines_1973}%
  \BibitemOpen
  \bibfield  {author} {\bibinfo {author} {\bibfnamefont {R.~S.}\ \bibnamefont
  {Hanni}}\ and\ \bibinfo {author} {\bibfnamefont {R.}~\bibnamefont
  {Ruffini}},\ }\href {\doibase 10.1103/PhysRevD.8.3259} {\bibfield  {journal}
  {\bibinfo  {journal} {Phys. Rev. D}\ }\textbf {\bibinfo {volume} {8}},\
  \bibinfo {pages} {3259} (\bibinfo {year} {1973})}\BibitemShut {NoStop}%
\bibitem [{\citenamefont {Wald}(1974)}]{wald_black_1974}%
  \BibitemOpen
  \bibfield  {author} {\bibinfo {author} {\bibfnamefont {R.~M.}\ \bibnamefont
  {Wald}},\ }\href {\doibase 10.1103/PhysRevD.10.1680} {\bibfield  {journal}
  {\bibinfo  {journal} {Phys. Rev. D}\ }\textbf {\bibinfo {volume} {10}},\
  \bibinfo {pages} {1680} (\bibinfo {year} {1974})}\BibitemShut {NoStop}%
\bibitem [{\citenamefont {Verlinde}(2011)}]{verlinde_origin_2011}%
  \BibitemOpen
  \bibfield  {author} {\bibinfo {author} {\bibfnamefont {E.}~\bibnamefont
  {Verlinde}},\ }\href {\doibase 10.1007/JHEP04(2011)029} {\bibfield  {journal}
  {\bibinfo  {journal} {J. High Energy Phys.}\ }\textbf {\bibinfo {volume}
  {2011}} (\bibinfo {year} {2011}),\ 10.1007/JHEP04(2011)029}\BibitemShut
  {NoStop}%
\bibitem [{\citenamefont {Kastor}\ \emph {et~al.}(2009)\citenamefont {Kastor},
  \citenamefont {Ray},\ and\ \citenamefont {Traschen}}]{kastor_enthalpy_2009}%
  \BibitemOpen
  \bibfield  {author} {\bibinfo {author} {\bibfnamefont {D.}~\bibnamefont
  {Kastor}}, \bibinfo {author} {\bibfnamefont {S.}~\bibnamefont {Ray}}, \ and\
  \bibinfo {author} {\bibfnamefont {J.}~\bibnamefont {Traschen}},\ }\href
  {\doibase 10.1088/0264-9381/26/19/195011} {\bibfield  {journal} {\bibinfo
  {journal} {Classical and Quantum Gravity}\ }\textbf {\bibinfo {volume}
  {26}},\ \bibinfo {pages} {195011} (\bibinfo {year} {2009})},\ \bibinfo {note}
  {arXiv: 0904.2765}\BibitemShut {NoStop}%
\bibitem [{\citenamefont
  {Dolan}(2011{\natexlab{a}})}]{dolan_cosmological_2011}%
  \BibitemOpen
  \bibfield  {author} {\bibinfo {author} {\bibfnamefont {B.~P.}\ \bibnamefont
  {Dolan}},\ }\href {\doibase 10.1088/0264-9381/28/12/125020} {\bibfield
  {journal} {\bibinfo  {journal} {Classical and Quantum Gravity}\ }\textbf
  {\bibinfo {volume} {28}},\ \bibinfo {pages} {125020} (\bibinfo {year}
  {2011}{\natexlab{a}})},\ \bibinfo {note} {arXiv: 1008.5023}\BibitemShut
  {NoStop}%
\bibitem [{\citenamefont {Cvetic}\ \emph {et~al.}(2011)\citenamefont {Cvetic},
  \citenamefont {Gibbons}, \citenamefont {Kubiznak},\ and\ \citenamefont
  {Pope}}]{cvetic_black_2011}%
  \BibitemOpen
  \bibfield  {author} {\bibinfo {author} {\bibfnamefont {M.}~\bibnamefont
  {Cvetic}}, \bibinfo {author} {\bibfnamefont {G.~W.}\ \bibnamefont {Gibbons}},
  \bibinfo {author} {\bibfnamefont {D.}~\bibnamefont {Kubiznak}}, \ and\
  \bibinfo {author} {\bibfnamefont {C.~N.}\ \bibnamefont {Pope}},\ }\href
  {\doibase 10.1103/PhysRevD.84.024037} {\bibfield  {journal} {\bibinfo
  {journal} {Phys. Rev. D}\ }\textbf {\bibinfo {volume} {84}} (\bibinfo {year}
  {2011}),\ 10.1103/PhysRevD.84.024037},\ \bibinfo {note} {arXiv:
  1012.2888}\BibitemShut {NoStop}%
\bibitem [{\citenamefont {Dolan}(2011{\natexlab{b}})}]{dolan_pressure_2011}%
  \BibitemOpen
  \bibfield  {author} {\bibinfo {author} {\bibfnamefont {B.~P.}\ \bibnamefont
  {Dolan}},\ }\href {\doibase 10.1088/0264-9381/28/23/235017} {\bibfield
  {journal} {\bibinfo  {journal} {Classical and Quantum Gravity}\ }\textbf
  {\bibinfo {volume} {28}},\ \bibinfo {pages} {235017} (\bibinfo {year}
  {2011}{\natexlab{b}})},\ \bibinfo {note} {arXiv: 1106.6260}\BibitemShut
  {NoStop}%
\bibitem [{\citenamefont {Kubiznak}\ and\ \citenamefont
  {Mann}(2012)}]{kubiznak_p-v_2012}%
  \BibitemOpen
  \bibfield  {author} {\bibinfo {author} {\bibfnamefont {D.}~\bibnamefont
  {Kubiznak}}\ and\ \bibinfo {author} {\bibfnamefont {R.~B.}\ \bibnamefont
  {Mann}},\ }\href {\doibase 10.1007/JHEP07(2012)033} {\bibfield  {journal}
  {\bibinfo  {journal} {J. High Energy Phys.}\ }\textbf {\bibinfo {volume}
  {2012}} (\bibinfo {year} {2012}),\ 10.1007/JHEP07(2012)033},\ \bibinfo {note}
  {arXiv: 1205.0559}\BibitemShut {NoStop}%
\bibitem [{\citenamefont {Bekenstein}(1973)}]{bekenstein_black_1973}%
  \BibitemOpen
  \bibfield  {author} {\bibinfo {author} {\bibfnamefont {J.~D.}\ \bibnamefont
  {Bekenstein}},\ }\href {\doibase 10.1103/PhysRevD.7.2333} {\bibfield
  {journal} {\bibinfo  {journal} {Phys. Rev. D}\ }\textbf {\bibinfo {volume}
  {7}},\ \bibinfo {pages} {2333} (\bibinfo {year} {1973})}\BibitemShut
  {NoStop}%
\bibitem [{\citenamefont {Hooft}(1993)}]{hooft_dimensional_1993}%
  \BibitemOpen
  \bibfield  {author} {\bibinfo {author} {\bibfnamefont {G.~t.}\ \bibnamefont
  {Hooft}},\ }\href {http://arxiv.org/abs/gr-qc/9310026} {\  (\bibinfo {year}
  {1993})},\ \bibinfo {note} {arXiv: gr-qc/9310026}\BibitemShut {NoStop}%
\bibitem [{\citenamefont {Susskind}(1995)}]{susskind_world_1995}%
  \BibitemOpen
  \bibfield  {author} {\bibinfo {author} {\bibfnamefont {L.}~\bibnamefont
  {Susskind}},\ }\href {\doibase 10.1063/1.531249} {\bibfield  {journal}
  {\bibinfo  {journal} {Journal of Mathematical Physics}\ }\textbf {\bibinfo
  {volume} {36}},\ \bibinfo {pages} {6377} (\bibinfo {year}
  {1995})}\BibitemShut {NoStop}%
\bibitem [{\citenamefont {Novikov}(1997)}]{novikov_black_1997}%
  \BibitemOpen
  \bibfield  {author} {\bibinfo {author} {\bibfnamefont {I.}~\bibnamefont
  {Novikov}},\ }in\ \href
  {http://link.springer.com/chapter/10.1007/3-540-31628-0_3} {\emph {\bibinfo
  {booktitle} {Stellar {Remnants}}}},\ \bibinfo {series and number} {\bibinfo
  {series} {Saas-{Fee} {Advanced} {Courses}}\ No.~\bibinfo {number} {25}}\
  (\bibinfo  {publisher} {Springer Berlin Heidelberg},\ \bibinfo {year}
  {1997})\ pp.\ \bibinfo {pages} {237--334}\BibitemShut {NoStop}%
\bibitem [{\citenamefont {Chang-Young}\ \emph {et~al.}(2011)\citenamefont
  {Chang-Young}, \citenamefont {Eune}, \citenamefont {Kimm},\ and\
  \citenamefont {Lee}}]{chang-young_schwarzschild-sitter_2011}%
  \BibitemOpen
  \bibfield  {author} {\bibinfo {author} {\bibfnamefont {E.}~\bibnamefont
  {Chang-Young}}, \bibinfo {author} {\bibfnamefont {M.}~\bibnamefont {Eune}},
  \bibinfo {author} {\bibfnamefont {K.}~\bibnamefont {Kimm}}, \ and\ \bibinfo
  {author} {\bibfnamefont {D.}~\bibnamefont {Lee}},\ }\href {\doibase
  10.1142/S0217732311036450} {\bibfield  {journal} {\bibinfo  {journal} {Modern
  Physics Letters A}\ }\textbf {\bibinfo {volume} {26}},\ \bibinfo {pages}
  {1975} (\bibinfo {year} {2011})},\ \bibinfo {note} {arXiv:
  1011.3960}\BibitemShut {NoStop}%
\bibitem [{\citenamefont {Jacobson}(1995)}]{jacobson_thermodynamics_1995}%
  \BibitemOpen
  \bibfield  {author} {\bibinfo {author} {\bibfnamefont {T.}~\bibnamefont
  {Jacobson}},\ }\href {\doibase 10.1103/PhysRevLett.75.1260} {\bibfield
  {journal} {\bibinfo  {journal} {Phys. Rev. Lett.}\ }\textbf {\bibinfo
  {volume} {75}},\ \bibinfo {pages} {1260} (\bibinfo {year}
  {1995})}\BibitemShut {NoStop}%
\bibitem [{\citenamefont {li}\ and\ \citenamefont
  {Zhao}(2000)}]{li_entropy_2000}%
  \BibitemOpen
  \bibfield  {author} {\bibinfo {author} {\bibfnamefont {X.}~\bibnamefont
  {li}}\ and\ \bibinfo {author} {\bibfnamefont {Z.}~\bibnamefont {Zhao}},\
  }\href {\doibase 10.1103/PhysRevD.62.104001} {\bibfield  {journal} {\bibinfo
  {journal} {Phys. Rev. D}\ }\textbf {\bibinfo {volume} {62}} (\bibinfo {year}
  {2000}),\ 10.1103/PhysRevD.62.104001}\BibitemShut {NoStop}%
\bibitem [{\citenamefont {li}\ and\ \citenamefont
  {Zhao}(2001)}]{li_black_2001}%
  \BibitemOpen
  \bibfield  {author} {\bibinfo {author} {\bibfnamefont {X.}~\bibnamefont
  {li}}\ and\ \bibinfo {author} {\bibfnamefont {Z.}~\bibnamefont {Zhao}},\
  }\href@noop {} {\bibfield  {journal} {\bibinfo  {journal} {Int. J. Theor.
  Phys.}\ ,\ \bibinfo {pages} {9}} (\bibinfo {year} {2001})}\BibitemShut
  {NoStop}%
\bibitem [{\citenamefont {Liu}\ and\ \citenamefont
  {Zhao}(2001)}]{liu_improved_2001}%
  \BibitemOpen
  \bibfield  {author} {\bibinfo {author} {\bibfnamefont {W.}~\bibnamefont
  {Liu}}\ and\ \bibinfo {author} {\bibfnamefont {Z.}~\bibnamefont {Zhao}},\
  }\href {\doibase 10.1088/0256-307X/18/2/353} {\bibfield  {journal} {\bibinfo
  {journal} {Chin. Phys. Letts.}\ }\textbf {\bibinfo {volume} {18}},\ \bibinfo
  {pages} {310} (\bibinfo {year} {2001})}\BibitemShut {NoStop}%
\bibitem [{\citenamefont {'t~Hooft}(1985)}]{t_hooft_quantum_1985}%
  \BibitemOpen
  \bibfield  {author} {\bibinfo {author} {\bibfnamefont {G.}~\bibnamefont
  {'t~Hooft}},\ }\href {\doibase 10.1016/0550-3213(85)90418-3} {\bibfield
  {journal} {\bibinfo  {journal} {Nuclear Physics B}\ }\textbf {\bibinfo
  {volume} {256}},\ \bibinfo {pages} {727} (\bibinfo {year}
  {1985})}\BibitemShut {NoStop}%
\end{thebibliography}%
%\begin{thebibliography}{0}
%%%% The number of zeroes here should correspond to the number of digits in the
%%%% number of bibliography entries here: if there are up to 9 entries, put one
%%%% zero; if there 10 up to 99 entries, put two zeroes; and so on.
%  
%  \bibitem{bib1}% 
%  \textsc{A.\,B.~Firstauthor}, 
%  \textsc{C.\,D.~Secondauthor}, and 
%  \textsc{E.~Lastauthor},
%  \jr{Abbreviatedjournalname} \textbf{volume}, page (year).
%  
%  \bibitem{bib2}% 
%  \textsc{X.~Ample} and  
%  \textsc{A.\,N.~Other},
%  \jr{Laser Phys. Rev.}   \textbf{1}, 111 (2050).
%  
%  \othercit % give this if it's not a properly cited journal with complete details
%  \bibitem{bib3}% 
%  \textsc{A.~Firstauthor}, 
%  \textsc{B.~Secondauthor}, and 
%  \textsc{C.~Thirdauthor},
%  The Title of the Book (Publisher, City, year), p.\,111.
%  
%  \othercit
%  \bibitem{bib4}% 
%  \textsc{A.~Firsteditor}, 
%  \textsc{B.~Secondeditor}, and 
%  \textsc{C.~Thirdeditor} (eds.),
%  The Title of the Edited Book (Wiley-VCH, Berlin, 2050), p.\,222.
%  
%  \othercit
%  \bibitem{bib5}% 
%  \textsc{D.~Contributor},
%  in:
%  The Title of the Edited Book,
%  edited by
%  A.~Firsteditor and B.~Secondeditor,
%  Title of the Series of Books [if any], volume number [if any] 
%  (Publisher, City, year), chap.\,1.
%  
%  \othercit
%  \bibitem{bib6}% 
%  \textsc{A.~Nother},
%  Proceedings of the 42nd Great Big Conference on Citation Formatting, Somewhere City,
%  Country, Year, Part A (Publisher, City, year),  pp.\,1--11.
%  
%\end{thebibliography}
\end{document}